\documentclass[twocolumn]{aastex63}

\submitjournal{ApJ}

\usepackage{comment}

\begin{document}

\author[0000-0002-8770-6764]{R\'eka K\"onyves-T\'oth}
\affiliation{Konkoly Observatory, Research Centre for Astronomy and Earth Sciences, Konkoly-Thege M. ut 15-17, Budapest, 1121 Hungary}

\author{B. P. Thomas}
\affiliation{Department of Astronomy, University of Texas at Austin, 2515 Speedway, Stop C1400 Austin, TX, USA}

\author[0000-0001-8764-7832]{J. Vink\'o}
\affiliation{Konkoly Observatory, Research Centre for Astronomy and Earth Sciences, Konkoly-Thege M. ut 15-17, Budapest, 1121 Hungary}
\affiliation{Department of Optics and Quantum Electronics, University of Szeged, D{\'o}m t\'er 9, Szeged, 6720 Hungary}
\affiliation{Department of Astronomy, University of Texas at Austin, 2515 Speedway, Stop C1400 Austin, TX, USA}

\author{J. C. Wheeler}
\affiliation{Department of Astronomy, University of Texas at Austin, 2515 Speedway, Stop C1400 Austin, TX, USA}

\shorttitle{Spectroscopy of SLSN 2019neq}
\shortauthors{K\"onyves-T\'oth et al.}

\correspondingauthor{R\'eka K\"onyves-T\'oth}
\email{konyvestoth.reka@csfk.mta.hu}

\graphicspath{{./}{}}

\title{Comparative Spectral Analysis of the Superluminous Supernova 2019neq }

\begin{abstract}

We present a detailed spectroscopic analysis of the recently  discovered fast evolving Type I superluminous supernova (SLSN-I), SN~2019neq (at redshift $z$~=~0.1059) comparing it to the well-studied slow evolving SLSN-I, SN~2010kd ($z$~=~0.101). Our investigation concentrates on optical spectra taken during the photospheric phase. The observations of SN~2019neq were carried out with the  10m  Hobby-Eberly  Telescope (HET)  Low  Resolution Spectrograph-2 (LRS2) at McDonald Observatory. We apply the {\tt SYN++} code to model the spectra taken at -4 days, +5 days and +29 days from maximum light. We examine the chemical evolution and ejecta composition of the SLSN by identifying the elements and ionization states in its spectra. Our analysis confirms that SN~2019neq is a fast evolving SLSN-I. We derive the number density of each ionization state at the epoch of the three observations. Finally, we give constraints on the lower limit of the ejecta mass and find a hint for a possible relation between the evolution timescale and the ejected mass of SLSNe-I.

\end{abstract}


\section{Introduction} \label{sec:intro}

A new class of supernovae, the so-called superluminous supernovae (SLSNe), was discovered and studied in the past two decades. At first, SLSNe were identified by their intrinsically high absolute magnitudes \citep[$\leq$-21 mag in all bands of the optical wavelengths;][]{galyam12}. Subsequently, the increasing amount of observational data led to a new classification scheme based on the spectroscopic properties of these events \citep{inserra19}. Like classical supernova types, SLSNe are divided into two main groups: the hydrogen-rich Type II SLSNe (SLSNe-II), and the hydrogen-poor SLSNe-I classes \citep{branch17}.
 SLSNe-II are separated into the following subclasses: SLSNe-IIn, with a luminosity evolution powered by an interaction with a massive circumstellar medium (CSM) (e.g. SN~2006gy; \citet{smith07}, and normal SLSNe-II, ostensibly without interaction (e.g. SN~2013hx; \citealt{inserra18}). The former have spectroscopic properties similar to traditional Type IIn SNe \citep{branch17}.
SLSNe-I have also been divided into two subgroups \citep{inserra18}: the fast evolving SLSN-I have an average light curve rise time of $\sim 28$ days (Fast SLSNe-I, e.g. SN~2015bn;  \citealt{nicholl16,nicholl18}), and the slow evolving SLSNe-I with rise time of $\sim 52$ days (Slow SLSNe-I e.g. SN~2011ke; \citealt{inserra13,quimby18}). \citet{inserra18} found that Fast SLSNe-I also exhibit high expansion velocities ($v \gtrsim 12000$ km~s$^{-1}$) and large velocity gradients, contrary to Slow SLSNe-I that are characterized by slower expansion velocities ($v \lesssim 12000$ km~s$^{-1}$) and negligible velocity gradients.   

In this paper, we present a comparative spectroscopic study of the recently discovered \citep{perley19}, and relatively close SN~2019neq ($z$~=~0.1059), which belongs to the fast-evolving SLSNe-I \citep{ben}, with the well-observed Slow SLSN-I, SN~2010kd \citep[$z$~=~0.101,][]{vinko10, kumar19}. Our main goal is to explore the differences between the two groups of SLSNe-I apart from the dissimilarity  in their light curve evolution timescale, and the differences in their velocity evolution \citep{inserra18}. This is a crucial question, because the differences in the spectrum may imply different ejecta. In a companion paper \citet{ben} explores the rate of spectroscopic evolution and the velocity gradient of SN~2019neq, finding that SN~2019neq is a SLSN-I with 
high velocity gradient and fast spectroscopic evolution. 

The slowly evolving SN~2010kd is a good comparison object to SN~2019neq in terms of chemical composition and spectral evolution, since the two SLSNe have similar redshifts. Recently \citet{kumar19} performed a detailed study of the photometric and spectroscopic properties of SN~2010kd.

The observations of the SLSNe that are examined herein, were carried out  using  the  10m  Hobby-Eberly  Telescope Low Resolution Spectrograph (LRS) and the Low Resolution Spectrograph-2 (LRS2) at McDonald Observatory. The  description of these observations is elaborated in \citet{kumar19} for SN~2010kd, and in \citet{ben} for SN~2019neq. The basic data of these two SLSNe can be found in Table \ref{tab:basic}.

In Section \ref{sec:models} we present detailed spectroscopic modeling of SN~2019neq with the code {\tt SYN++} \citep{thomas11}. We use 3 spectra of the object taken at different epochs during the photospheric phase. We identify the chemical composition of the ejecta and reveal its spectroscopic evolution. 

The early-phase spectra of SLSNe are dominated by the W-like feature around $\sim 4500$ \AA\ that is widely accepted as being due to O II, although other suggestions also exist \citep[e.g.][]{quimby07}. \citet{quimby18} presented an in-depth analysis of this region using {\tt SYN++} found evidence supporting the O II hypothesis, also favored by \citet{mazzali16}, for most of their sample SLSNe-I. 
We re-examine this issue here, in the case of SN~2019neq, because this SLSN showed relatively hot and fast-expanding ejecta during the early phases. Under such circumstances the presence/absence of ionization states and their associated features can be temperature constrained.   

In Section \ref{sec:disc}, first, we discuss the classification of this object (Section \ref{clas}), then infer the number densities of the identified elements for each detected ionization state (Section~\ref{infer}).
In Section \ref{comp} we compare the spectral features and the spectroscopic evolution of SN~2010kd and SN~2019neq.  We estimate the total mass ejected during the explosion in Section \ref{mass}.
Finally, in Section \ref{sec:concl}, we summarise our conclusions.

\section{Spectrum modeling} \label{sec:models}

\begin{table*}
\caption{Basic data of the studied SLSNe.}
\label{tab:basic}
\begin{center}
\begin{tabular}{lcccccccc}
\hline
\hline
SN & R.A. & Dec. & Discovery & Explosion & $t_{\rm max}$ & $z$ & $E(B-V)$ & Reference \\
& & & & (MJD) & (MJD) & & & \\
\hline 
SN~2010kd & 12:08:01 & +49:13:31 & 2010-11-14$^1$ &55499.5 &55552.0 & 0.1010 & 0.0197 & \citet{vinko10}\\
SN~2019neq & 17:54:26 & +47:25:40 & 2019-08-10$^2$ & 58700.0 & 58731.0 & 0.1059\tablenotemark{a} & 0.0330 & \citet{perley19}\\
\hline
\end{tabular}
\tablenotetext{a}{Calculated from narrow H$\alpha$ emission from the host.}
\end{center}
\end{table*}

\begin{figure*}
\begin{center}
\includegraphics[]{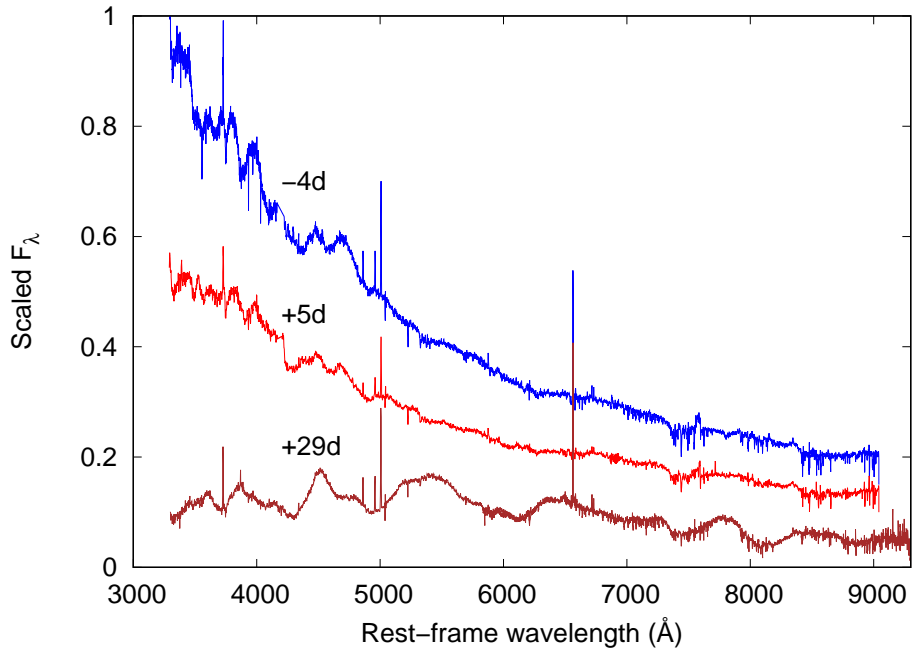}
\caption{The observed spectra of SN~2019neq at phase -4 days, +5 days and +29 days from maximum.}
\label{fig:spectra}
\end{center}
\end{figure*}

The spectra of SN~2019neq considered in this paper can be seen in Figure \ref{fig:spectra}.
These spectra were corrected for redshift and interstellar extinction before plotting. The  phase of the observed spectra were computed relative to the epoch of maximum light as determined by \citet{ben} to be 2019-09-05 (MJD 58731), allowing for the time dilation due to redshift.

\begin{figure}
\begin{center}
\includegraphics[width=8.5cm]{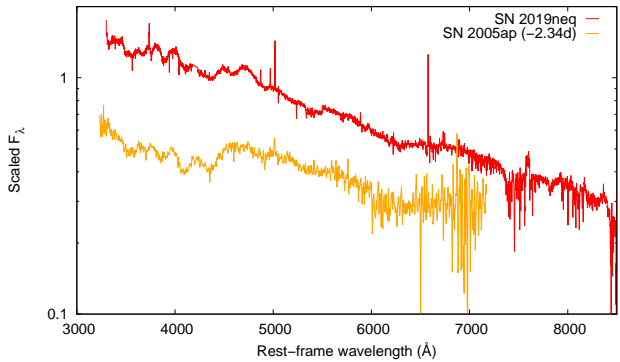}
\caption{ Comparison of the premaximum spectra of  SN~2019neq -4d (red) and SN~2005ap -2d (orange). The general similarity of the two spectra is apparent. }
\label{fig:neq_vs_ap}
\end{center}
\end{figure}

In Figure \ref{fig:neq_vs_ap}, the first spectrum of SN~2019neq taken at phase -4d (4 days before maximum light) is compared to a spectrum of SN~2005ap taken at similar phase (-2d) \citep{quimby07}. The similarity of the two spectra is apparent, as  was  noted by \citep{perley19}. Beyond the spectral similarity to the premaximum spectrum of SN 2005ap, the chemical composition of SN~2019neq, especially the lack of H and He features, 
provides significant evidence that shows that SN 2019neq is a SLSN-I near maximum light \citep{atel}.

To model the available photospheric phase spectra, we used the {\tt SYN++} \citep{thomas11} code, which is the revised and improved version of the FORTRAN code {\tt SYNOW} \citep{fisher99,hatano99}. 

In {\tt SYN++}, there are some global parameters referring to the whole model spectrum, and local parameters to fit the lines of the individual elements.
The global parameters are the following: 
\begin{itemize}
    \item {{$a_0$}: a constant multiplier to the whole model spectrum}
    \item {{$v_{\rm phot}$}: velocity at the photosphere}
    \item {{$T_{\rm phot}$}: temperature at the photosphere}
\end{itemize} 
The local parameters are:
\begin{itemize}
    \item {{$\tau$}: optical depth for the reference line of each ion}
    \item {{$v_{\rm min}$}: the inner velocity of the line forming region}
    \item {{$v_{\rm max}$}: the outer velocity of the line forming region}
    \item {{$\sigma$}: scale height of the optical depth above the photosphere in km s$^{-1}$. This parameter is responsible for the width of the spectral features, that is roughly related to the width of the line-forming region in the atmosphere. Larger $\sigma$ parameter implies a broader feature.}
    \item{{$T_{\rm exc}$:}  excitation temperature of each element, assuming LTE. Different ions may have different $T_{\rm exc}$ parameters, mimicking NLTE conditions. }
\end{itemize}

It can be  seen in Figure \ref{fig:spectra} that during the photospheric phase the spectra of SN~2019neq were dominated by a hot, blue continuum with strong, overlapping P Cygni features, even though the presence of emission lines due to NLTE effects cannot be ruled out. Since there is no single, unblended feature in these spectra, a spectrum synthesis code is necessary to determine  the chemical composition of the ejecta reliably, even with the assumption of LTE \citep{branch17}.

To examine the evolution of the temperature and photospheric velocity of SN~2019neq, and identify these firm P Cygni lines, we modelled the spectra taken at 3 epochs: -4, +5, and +29 rest-frame days relative to maximum light. The global parameters for the best-fit models are collected in Table 
 \ref{tab:sn19neq_globparams}, while the list of the local parameters for each ion can be found in the tables in the Appendix.

\begin{table}
\caption{Best-fit global parameters of the  SYN++ photospheric phase models of SN~2019neq.}
\label{tab:sn19neq_globparams}
\centering
\begin{tabular}{ccccc}
\hline 
MJD & Phase &  $a_0$& $v_{\rm phot}$ & $T_{\rm phot}$   \\
(days) & (days) &  & (km~s$^{-1}$) & ($10^3$ K) \\
\hline
\hline
58727 & -4 & 0.24 & 21 000 & 15.0 \\
58737 &  5 & 0.13 & 21 000 & 12.0 \\
58763 & 29 & 0.14 & 12 000 & 6.0 \\
\hline
\end{tabular}
\end{table}

\begin{figure*}
\begin{center}
\includegraphics[width=8.5cm]{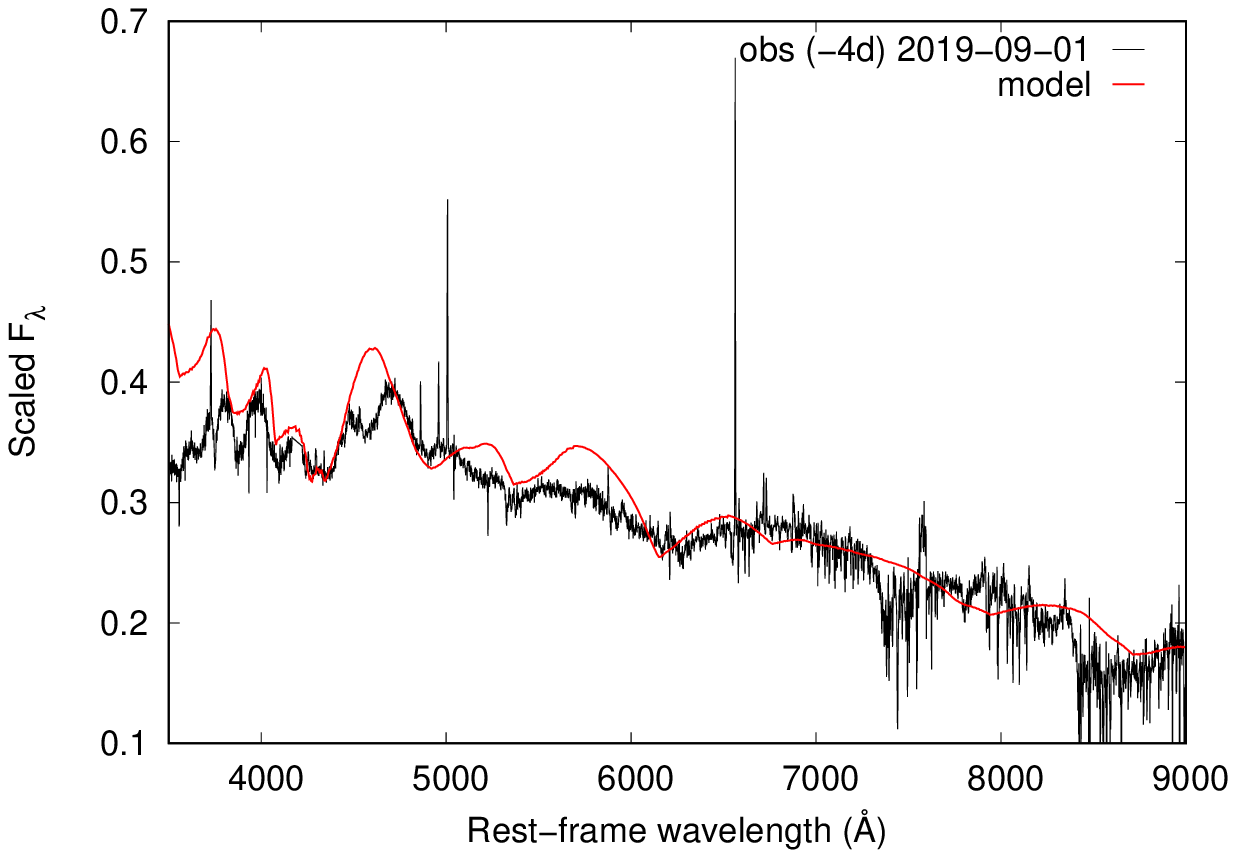}
\includegraphics[width=8.5cm]{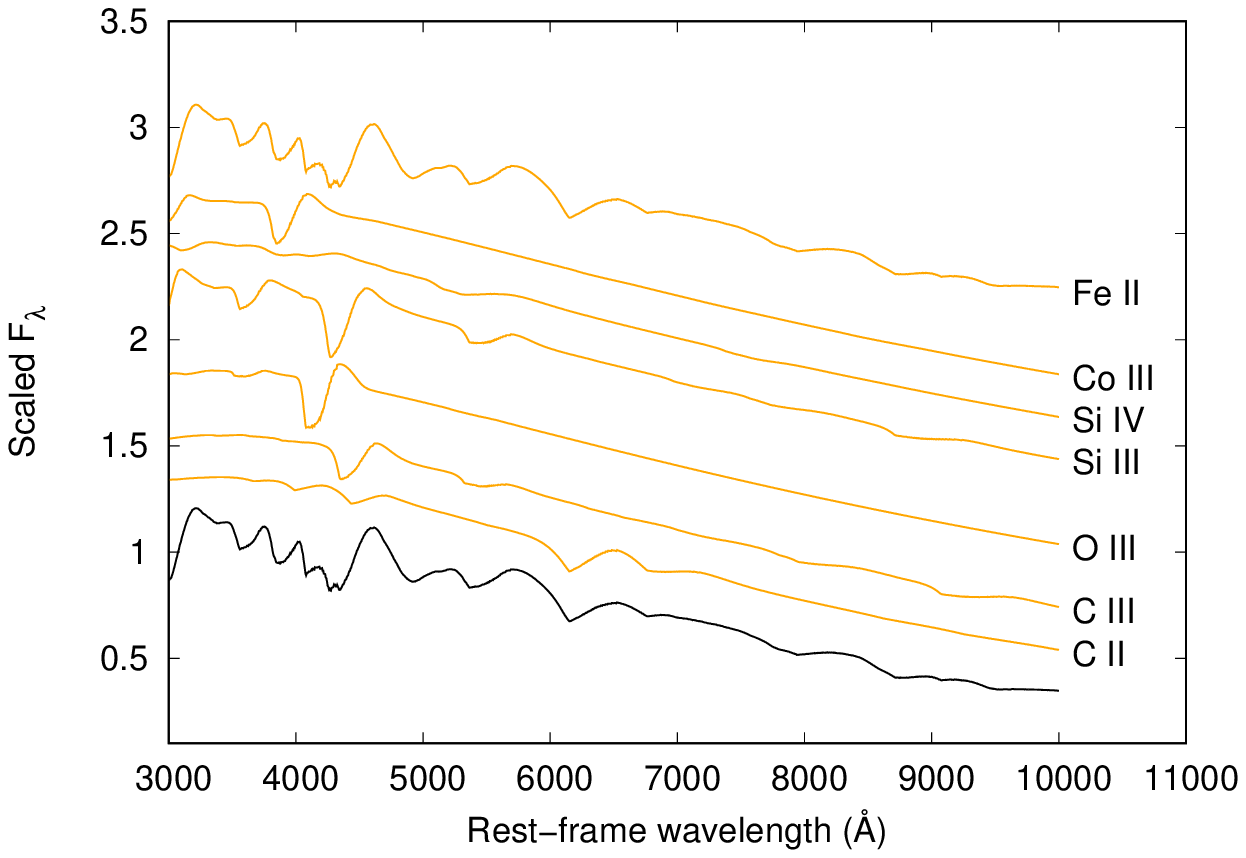}
\caption{Left panel: The observed (black line) spectrum of SN~2019neq at phase -4d (2019-09-01), plotted together with the best-fit model obtained with {\tt SYN++} (red line). On the vertical axis $\lambda^2 \cdot F_\lambda$ is plotted. Right panel: Single ion contributions (orange lines) to the overall model spectrum (black line).}
\label{fig:sn19neq_20190901}
\end{center}
\end{figure*}

The observed spectrum and its best-fit model for the first epoch (at phase -4d) can be seen in Figure \ref{fig:sn19neq_20190901}. The observed spectrum contains strong, narrow H$\alpha ~\lambda6562.8$ and forbidden [O III] $\lambda\lambda\lambda$4932, 4960, 5008  lines due to the host galaxy. The redshift of SN~2019neq was calculated by fitting a Gaussian profile to the narrow H$\alpha$ feature, resulting in $z~=~0.105942 \pm 0.000006$, as presented in Table \ref{tab:basic}.

\begin{figure*}
\begin{center}
\includegraphics[width=8.5cm]{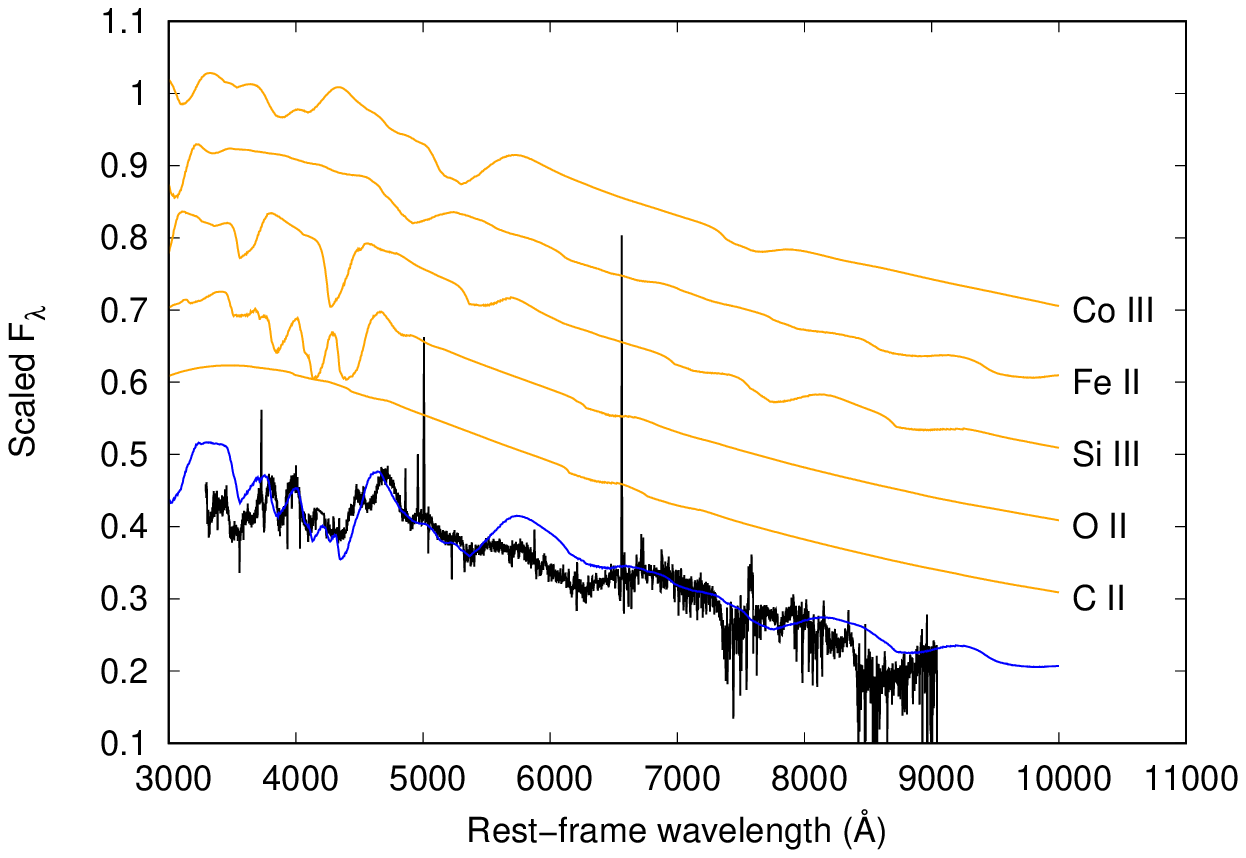}
\includegraphics[width=8.5cm]{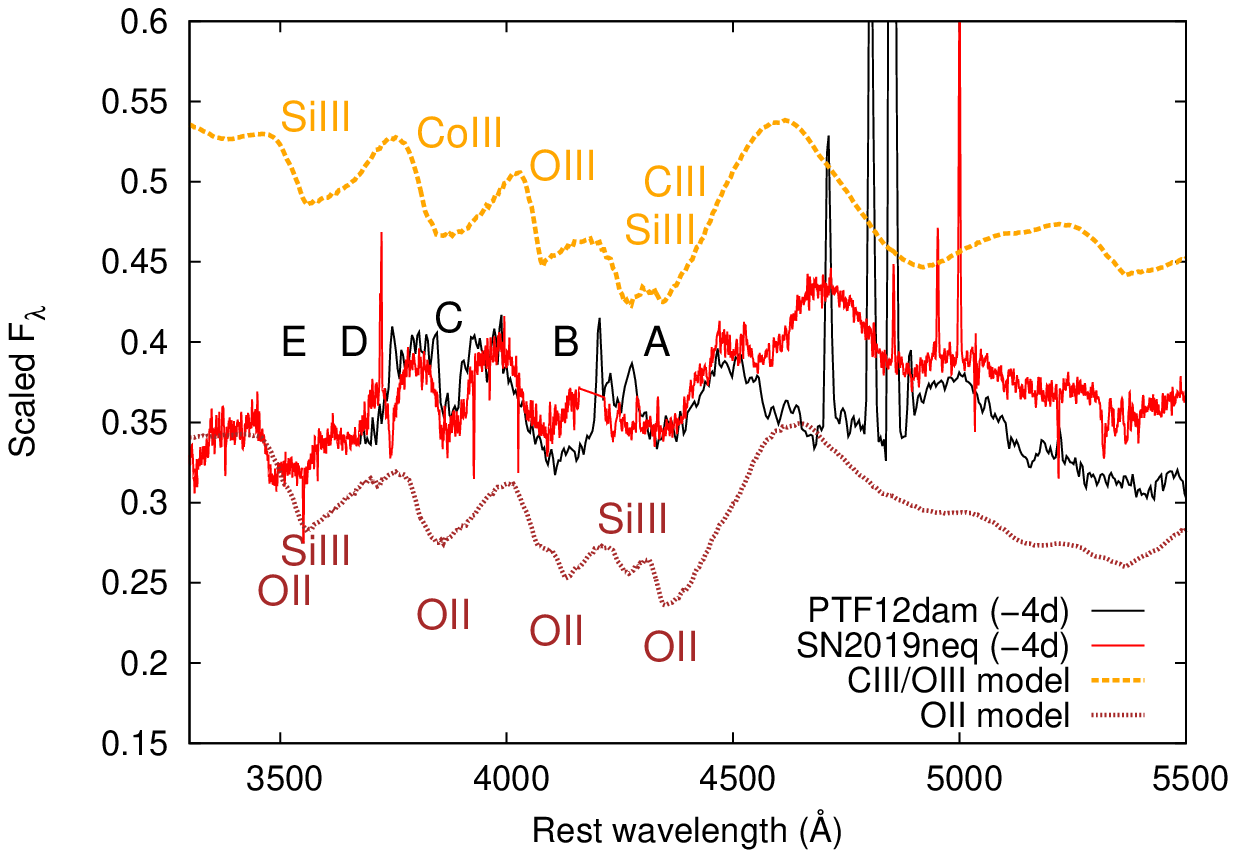}
\caption{Left panel: the alternative model of the -4d spectrum of SN~2019neq, where the W-shaped feature between 4000 and 5000 \AA\ is fitted with C II and O II instead of C III and O III. Right panel: Comparison of the two models with the -4d spectrum of SN~2019neq (red) and PTF12dam (black) taken at similar phase. The models have been shifted vertically for clarity, and all spectra have been flattened for continuum. Ion identifications are from {\tt SYN++}.}
\label{fig:sn19neq_20190901_alter}
\end{center}
\end{figure*}

At this early phase, the photospheric temperature of the {\tt SYN++} model is  15000 K, while the expansion velocity also seems to be very  high, 21000 km~s$^{-1}$, compared to normal Type-Ia or core collapse SNe. This photospheric velocity value is similar to the frequently identified high-velocity component in the pre-maximum spectra of Type Ia SNe \citep[e.g.][]{mulligan19,silver15}.

 Our first model, plotted in Fig. \ref{fig:sn19neq_20190901} contains C II, C III, O III, Si III, Si IV, Co III, and Fe II lines (optical depths and other parameters are summarized in tables in the Appendix). In particular, the W-shaped feature appearing between 4000 and 5000 \AA, which appears somewhat weaker in SN~2019neq than in other SLSNe-I shown in \citet{quimby18}, can be fitted with the combination of C III, O III, Si III and Co III multiplets, similar to the results by \citet{quimby07}. 

Alternatively, the spectrum can also be fitted by using O II and C II instead of O III and C III, as shown in the left panel in Fig. \ref{fig:sn19neq_20190901_alter}. The comparison of these two models is plotted in the right panel, where an observed (yet unpublished) HET spectrum of PTF12dam (black line) taken at a similar rest-frame phase as our first SN 2019neq spectrum (red line) is also plotted, together with the two {\tt SYN++} models. The labels A, B, C, D and E mark the same features as in \citet{quimby18} (see their Fig.13). It is seen that the two simple {\tt SYN++} models can explain the appearance of these features almost equally well (at least both of them are consistent with the observed spectrum), and neither of them accounts for the pseudo-emission around $\sim 4700$ \AA\ that is present in SN~2019neq and absent in PTF12dam. 

Even though the ``O II-model'' provides a more elegant explanation for 4 out of 5 observed features with a single ion, the validity of the ``O III-model'' cannot be ruled out in the hot ejecta of SN 2019neq. According to \citet{hatano99}, the optical depths for O II and C II expected in an atmosphere having $T~\sim$~15000 K are the same as for O III and C III, thus, it is possible that doubly-ionized ions also play a role in forming the spectrum of SN 2019neq between 3500 and 4500 \AA\ .

\begin{figure*}
\begin{center}
\includegraphics[width=8.5cm]{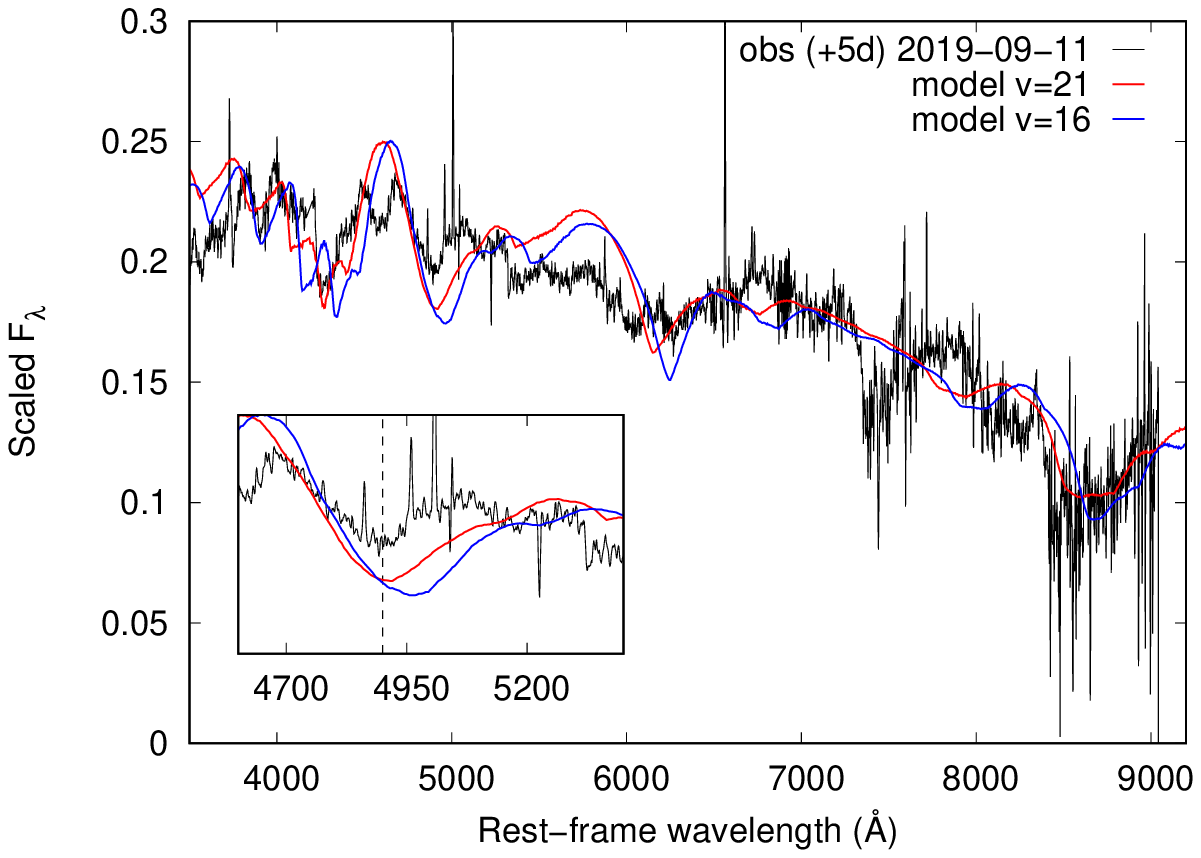}
\includegraphics[width=8.5cm]{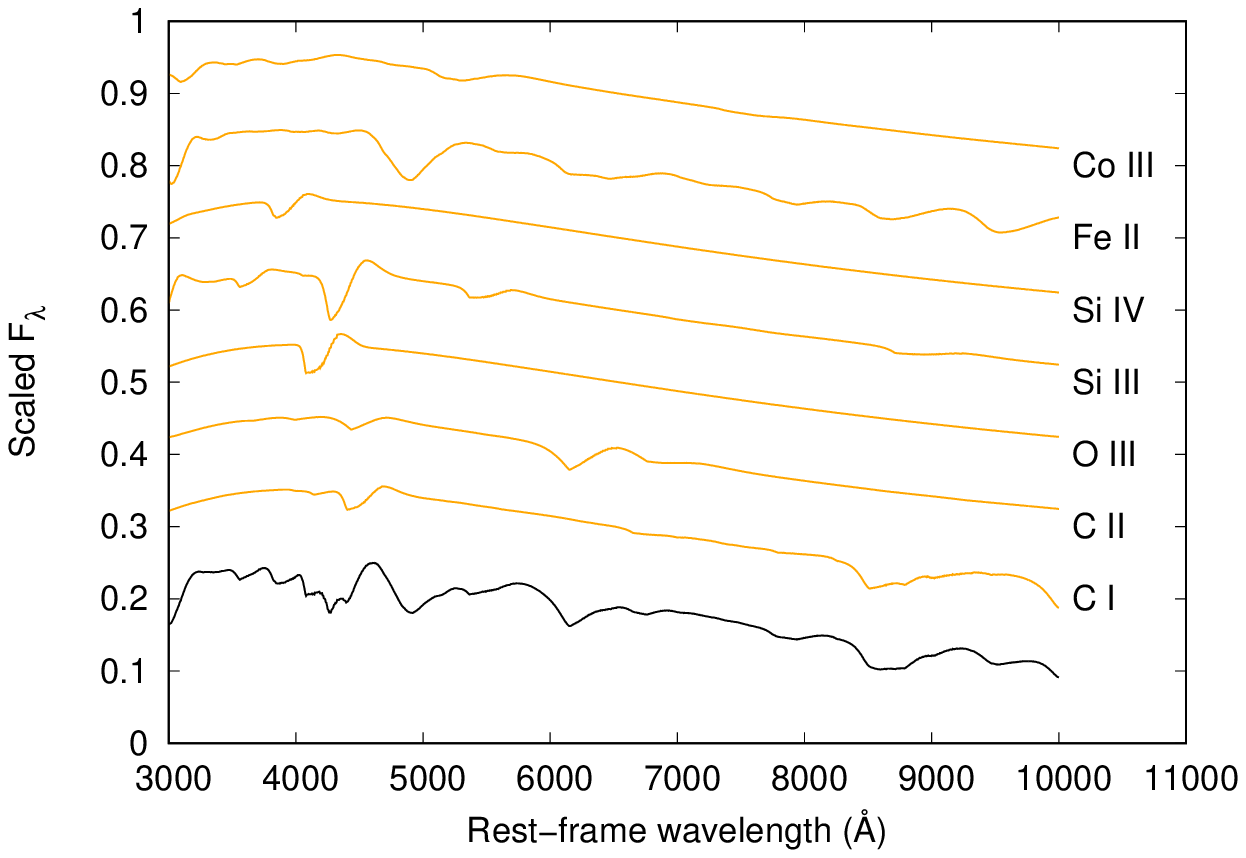}
\caption{The observed and modelled spectra of SN~2019neq at +5d phase
(2019-09-11). In the left panel the blue line shows a model with $v_{\rm phot}$~=~16000 km~s$^{-1}$, and the red line denotes the best-fit model with  $v_{\rm phot}$~=~21000 km~s$^{-1}$. The inset zooms in on the Fe II $\lambda$5169 feature. The absorption minimum of the red model is much closer to the observed minimum of this line, as indicated by the dashed vertical line in the inset.
The color coding of the right panel is the same as in Figure \ref{fig:sn19neq_20190901}.}
\label{fig:sn19neq_20190911}
\end{center}
\end{figure*}

The second spectrum of SN~2019neq was taken at +5d after maximum light. Figure 
\ref{fig:sn19neq_20190911} shows the best-fit {\tt SYN++} model to this spectrum, in which the photospheric temperature decreased to 12000 K, and the lines of C I also appeared.  In the left panel of Fig. \ref{fig:sn19neq_20190911}, two alternative models can be seen, having  the same local and global parameters but different photospheric velocities.
The ambiguity of $v_{\rm phot}$ is caused by the identification of the features around 5000 \AA\ thought to be due to Fe II. If we assigned the minimum of the observed feature (shown by the dashed vertical line in the inset of the left panel of Figure \ref{fig:sn19neq_20190911}) to the Fe II $\lambda$5169 transition, which is a strong observed Fe II line in Type II SNe, then the photospheric velocity would be 16000 km~s$^{-1}$. Accordingly, our first model for this spectrum was built with $v_{\rm phot}$ = 16000 km~s$^{-1}$ (plotted with blue in the left panel of Figure \ref{fig:sn19neq_20190911}), then a second model was also developed using the criterion that the absorption minima of all identified features are fit optimally. It was found that such a model has $v_{phot} \sim$ 21000 km~s$^{-1}$, which is shown by the red line in the left panel of Figure \ref{fig:sn19neq_20190911}.
It is seen that the model with a higher photospheric velocity matches the data more accurately than the slower model.
The right panel of Figure \ref{fig:sn19neq_20190911} presents the ion contributions to the best-fit model.

\begin{figure}
\begin{center}
\includegraphics[width=8.5cm]{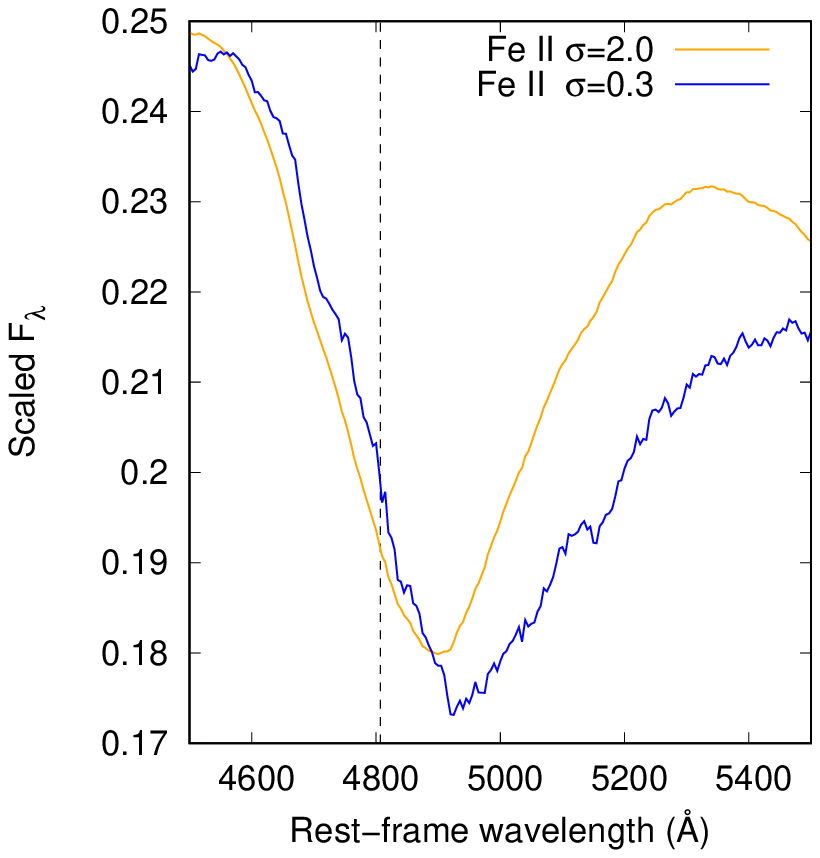}
\caption{{\tt SYN++} models of the Fe II  $\lambda$5169  feature, assuming $v_{phot} = 16000$ km s$^{-1}$. Different colors refer to different values of the $\sigma$ parameter, and the dashed line shows the  wavelength of the suspected absorption minimum of Fe II $\lambda$5169, Doppler-shifted to $v_{phot}~=~$16000 km~s$^{-1}$. The wavelengths of the minima of the two model spectra are different from the position of the vertical line, suggesting that this broad feature is not due to a single line of Fe II $\lambda$5169 .}
\label{fig:sn19neq_20190911_feii}
\end{center}
\end{figure}

To explore the cause of this inconsistency, we modelled the  Fe II lines with different values of the  $\sigma$ parameter in the vicinity of 5000 \AA\ , as can be seen in Figure \ref{fig:sn19neq_20190911_feii}. The orange line denotes $\sigma$~=~2000 km~s$^{-1}$, utilized in the  model having $v_{\rm phot}$~=~16000  km~s$^{-1}$, which is plotted together with another model having $\sigma$~=~300 km~s$^{-1}$ (blue line). The dashed vertical line shows the supposed wavelength of the Fe II $\lambda$5169 absorption minimum corresponding to $v_{\rm phot}$~=~16000  km~s$^{-1}$. It is seen that the feature assumed to be a strong Fe II $\lambda$5169 absorption line, is actually a blend of many weak features. The small humps on the blue curve correspond to these individual Fe II transitions, which become blended with each other on the orange curve when the widths of the features are broader (indicated by the higher $\sigma$ parameter).
It is clear that the Doppler-shifted position of the Fe II $\lambda$5169  (dashed vertical line)  differs from the wavelengths of the minima of the two model spectra. It is concluded that the broad observed feature around 5000 \AA\ cannot be interpreted simply as due to Fe II $\lambda$5169, and modelling the whole spectrum is necessary to reveal the true photospheric velocity. Since the model having 
$v_{\rm phot}~=~21000$ km s$^{-1}$ describes the data better than the model with
$v_{\rm phot}$~=~16000 km $s^{-1}$, we adopted the former value as the photospheric velocity of the best-fit model to the +5d spectrum.

\begin{figure*}
\begin{center}
\includegraphics[width=8.5cm]{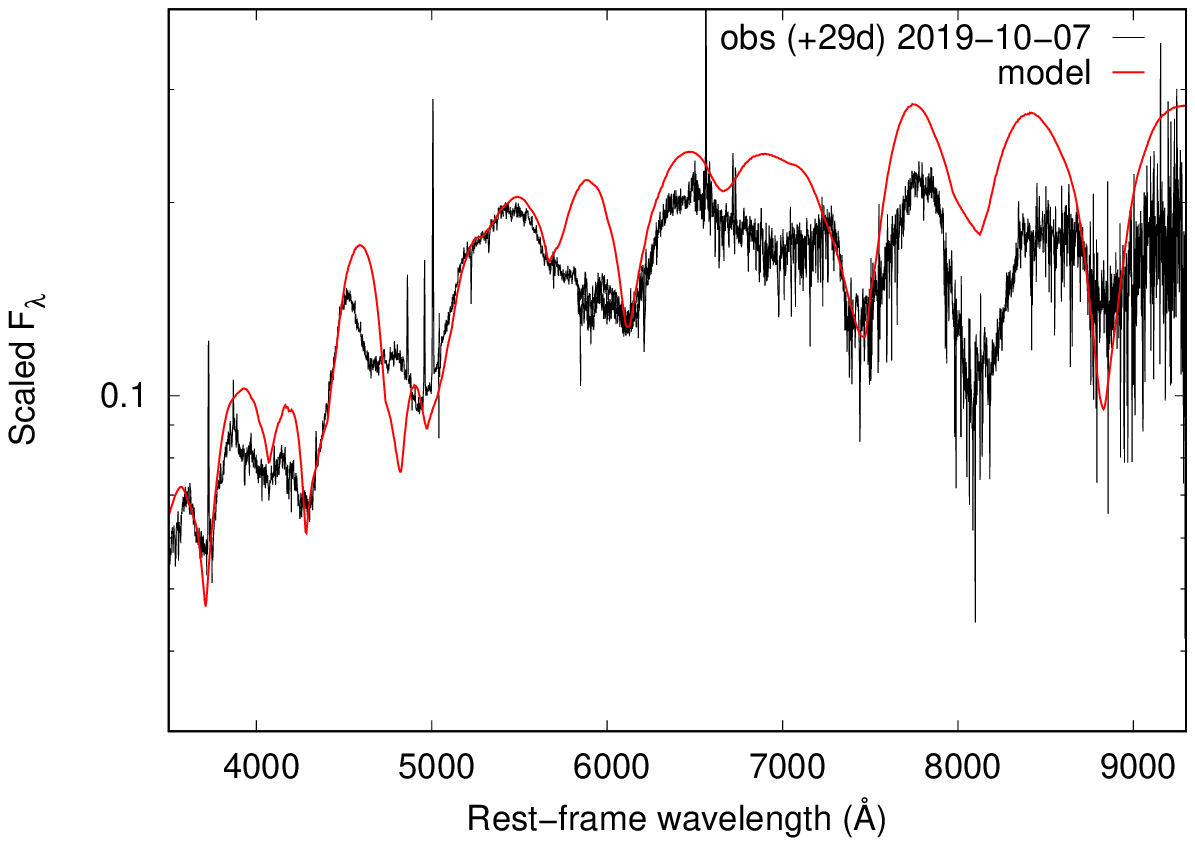}
\includegraphics[width=8.5cm]{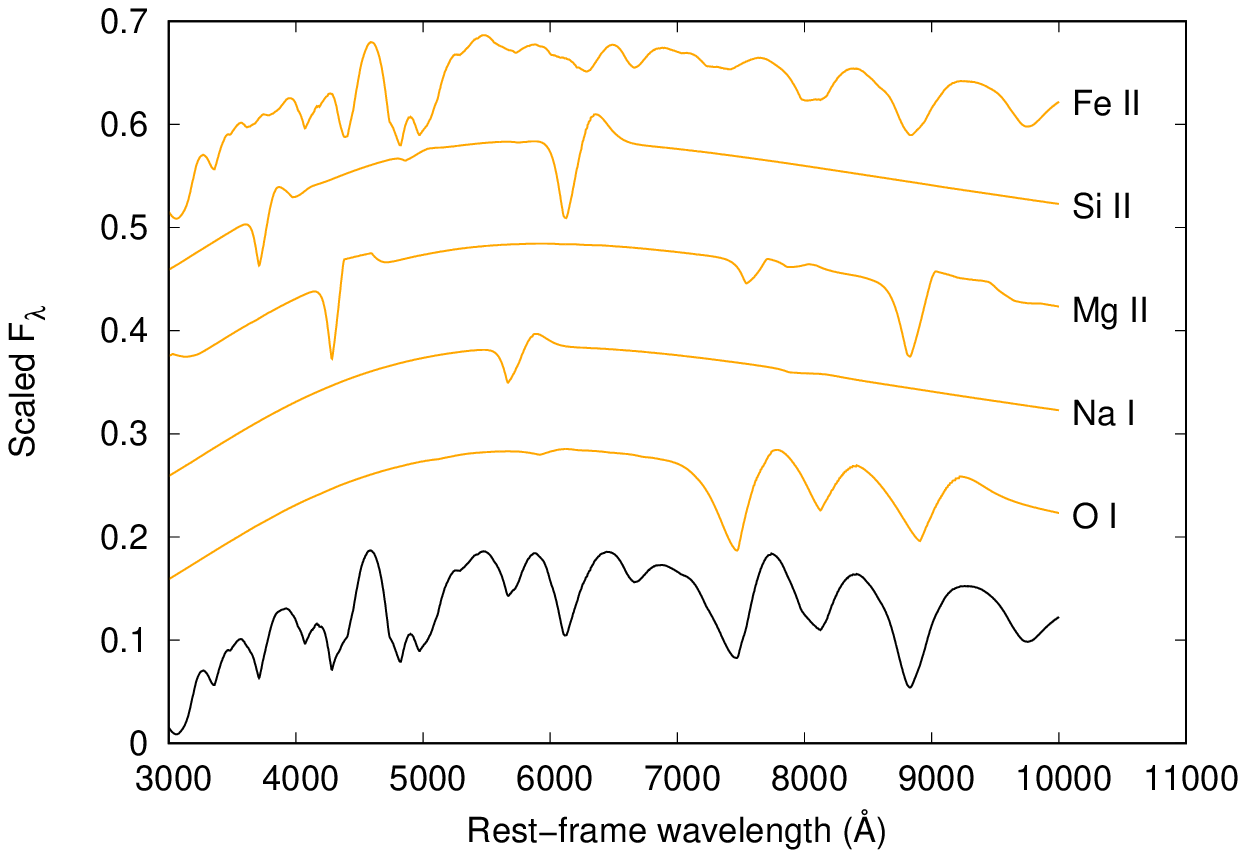}
\caption{{\tt SYN++} modeling of the +29d phase (2019-10-07) spectrum of SN~2019neq, with the same color coding as in Fig. \ref{fig:sn19neq_20190901}.} 
\label{fig:sn19neq_20191007}
\end{center}
\end{figure*}

Based on the spectrum taken at $\sim 1$ month after maximum, \citet{atel2} reported that 
SN~2019neq exhibited very fast spectral evolution. This can also be seen in Fig. \ref{fig:sn19neq_20191007}, where the third spectrum, taken at +29d phase is plotted together with its best-fit {\tt SYN++} model.
The photospheric velocity decreased from  21000 km~s$^{-1}$ to
12000 km~s$^{-1}$, and the temperature at the photosphere diminished from 12000 K to 6000 K . In accord with the decreasing temperature, the low ionization elements began to dominate the highly excited ones. We identified the presence of O I, Na I, Mg II, Si II, and Fe II lines, as can be seen in Figure \ref{fig:sn19neq_20191007} (see also Table \ref{tab:sn19neq_globparams}). 

\section{Discussion} \label{sec:disc}

\subsection{The classification of SN~2019neq as a fast evolving SLSN-I}\label{clas}

\begin{figure}
\begin{center}
\includegraphics[width=8cm]{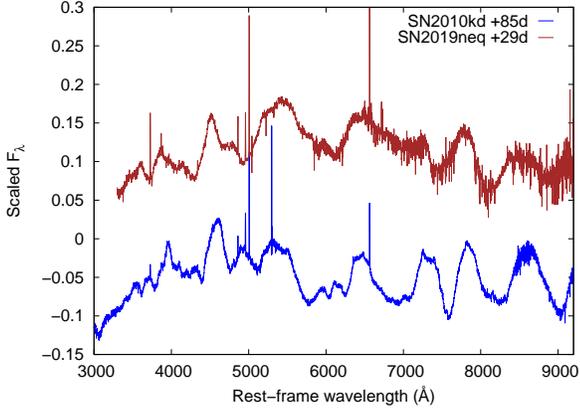}
\caption{Spectral comparison of SN~2010kd at +85d phase (blue line) and SN~2019neq at +29d (brown line). These spectra show similar features in spite of their different phases confirming the  fast spectral evolution of SN~2019neq.}
\label{fig:moses_vs_neq}
\end{center}
\end{figure}

In Figure \ref{fig:moses_vs_neq}, we compared the +29d spectrum of  SN~2019neq to that of SN~2010kd taken at +85d phase \citep{kumar19}. The features of the two spectra are quite similar, in spite of their different phases. SN~2010kd was a slowly evolving SLSN-I \citep{kumar19}. SN~2019neq reached the same physical stage at $\sim$30 days as SN~2010kd at +85d phase, illustrating the  fast spectral evolution of SN~2019neq (see also \citealt{ben}).
Note that in both spectra some nebular emission features (e.g. [O I] $\lambda \lambda$6300,6363; [Ca II] $\lambda \lambda$7291,7323) seem to start appearing, which suggest the dilution of the ejecta and strengthening of the NLTE conditions in the envelope. 

\subsection{Inferring the number density of the ionization states in the ejecta of SN~2019neq}\label{infer}

From the {\tt SYN++} model parameters listed in Tables \ref{tab:sn19neq_globparams} and \ref{tab:neq_lokparams},
we estimated the number and mass densities of the identified ions in  each spectrum, following \citet{hatano99}.

According to the Sobolev-approximation
\citep[e.g.][]{hatano99}, the optical depth of a P Cygni feature can be expressed as
\begin{equation}
    \tau~=~\left({{\pi e^2} \over {m_ec}}\right) f \lambda t n_l \left(1 - {{g_l n_u} \over {g_u n_l}}\right),
    \label{eq:logtau}
    \end{equation}
where   $n_u$ and $n_l$ refer to the number densities of the particular ion at the upper and lower levels of the transition, $g_u$ and $g_l$ are the statistical weights, $f$ is the oscillator strength, $t$ is the rest-frame time since explosion, $e$ and $m_e$ are the charge and the mass of an electron, and $c$ is the speed of the light.

The LTE conditions adopted by {\tt SYN++} imply that 
\begin{equation}
    {{n_u} \over {n_l}}~=~{{g_u} \over {g_l}} e^{- {{(E_u-E_l)} \over {k T}} }.
    \label{eq:boltz}
    \end{equation}
Here, $E_u$ and $E_l$ are the energies of the upper and lower levels, and $T$ is the excitation temperature.
   
From  Equations \ref{eq:logtau} and \ref{eq:boltz}, 
we can calculate the optical depth as
\begin{equation}
    \tau~=~0.026f \lambda_\mu t_d n_l \left(  1 - e^{ - {{hc} \over {\lambda kT}} }  \right),
    \label{eq:taunum}
    \end{equation}
where $\lambda_\mu$ is the wavelength of a particular feature in $\mu$m, and $t_d$ is the number of rest-frame days from explosion. 

The value of $n_l$ can then be expressed as 
\begin{equation}
    n_l~=~ { {\tau} \over { 0.026f \lambda_\mu t_d \left(  1 - e^{ - {{hc} \over {\lambda kT}} }  \right) } }~.
    \label{eq:nl}
\end{equation}

To get the full number density of an element, we can apply the alternative form of the Boltzmann formula (Eq. \ref{eq:boltz}):
\begin{equation}
   {{n_l} \over {N}} ~=~ {{g_l} \over {z(T)}} \cdot e^{ -{{\chi} \over {kT}}},
    \label{eq:altboltz}
\end{equation}
where $N$ denotes the full number density of an ion in cm$^{-3}$, $z(T)$ is the partition function, $\chi~=~E_l-E_0$  is the excitation potential of the lower level (in eV), and $T$ is the excitation temperature (in K). 

From  Equation \ref{eq:altboltz} the total number density ($N$) can be inferred as
\begin{equation}
   N~=~ {{n_l~ z(T)} \over {g_l}} \cdot e^{ {{5040 \over T} \cdot \chi}} ~,
    \label{eq:N}
\end{equation}
where $n_l$  is given by Eq. \ref{eq:nl}.

From the equations above, the density of each ionization state  (in g~cm$^{-3}$) can be calculated as the product of the full number density and the ion mass. 

In the case of SN~2019neq, the inferred $n_l$ and $N$ values, as well as the densities for each identified ionization state can be found in Table \ref{tab:koncentracio_out_neq} in the Appendix. The required data for these calculations can be seen in Table \ref{tab:koncentracio_input_neq} in the Appendix, and come from the following sources:  $\tau$, $T$ come from the  {\tt SYN++} model file, while the atomic data are collected from \citet{hatano99}, and the NIST (National Institute of Standards and Technology) Atomic Spectra Database \footnote{https://www.nist.gov/pml/atomic-spectra-database}. Note that the ion densities for O II and O III were omitted from Table \ref{tab:koncentracio_input_neq} and \ref{tab:koncentracio_out_neq}, because the reference lines for these ions are forbidden transitions  \citep[see Table 2 in][]{hatano99}. Since {\tt SYN++} calculates the occupation numbers of the different atomic levels assuming LTE, we found that this leads to very high uncertainties  
in the inferred number densities when the reference lines are forbidden, probably due to the breakdown of the LTE assumption for such transitions.

We conclude that the identified ions and their number densities belonging to the first and second epochs ($-4$ and +5 days, respectively) are quite similar. On the contrary, the third spectrum (taken at +29 days) contains a variety of different species, thus, the calculated  densities are also different. This may suggest that the inner region of the ejecta, revealed by the spectra at later epochs, is richer in heavier elements than the upper parts of the atmosphere. To explore the chemical evolution of SN~2019neq in more detail, we need to follow-up the object with further spectroscopic observations in the future. 

\subsection{Comparing the spectral evolution with the Slow SLSN-I SN~2010kd}\label{comp}

\begin{figure*}
\begin{center}
\includegraphics[width=8cm]{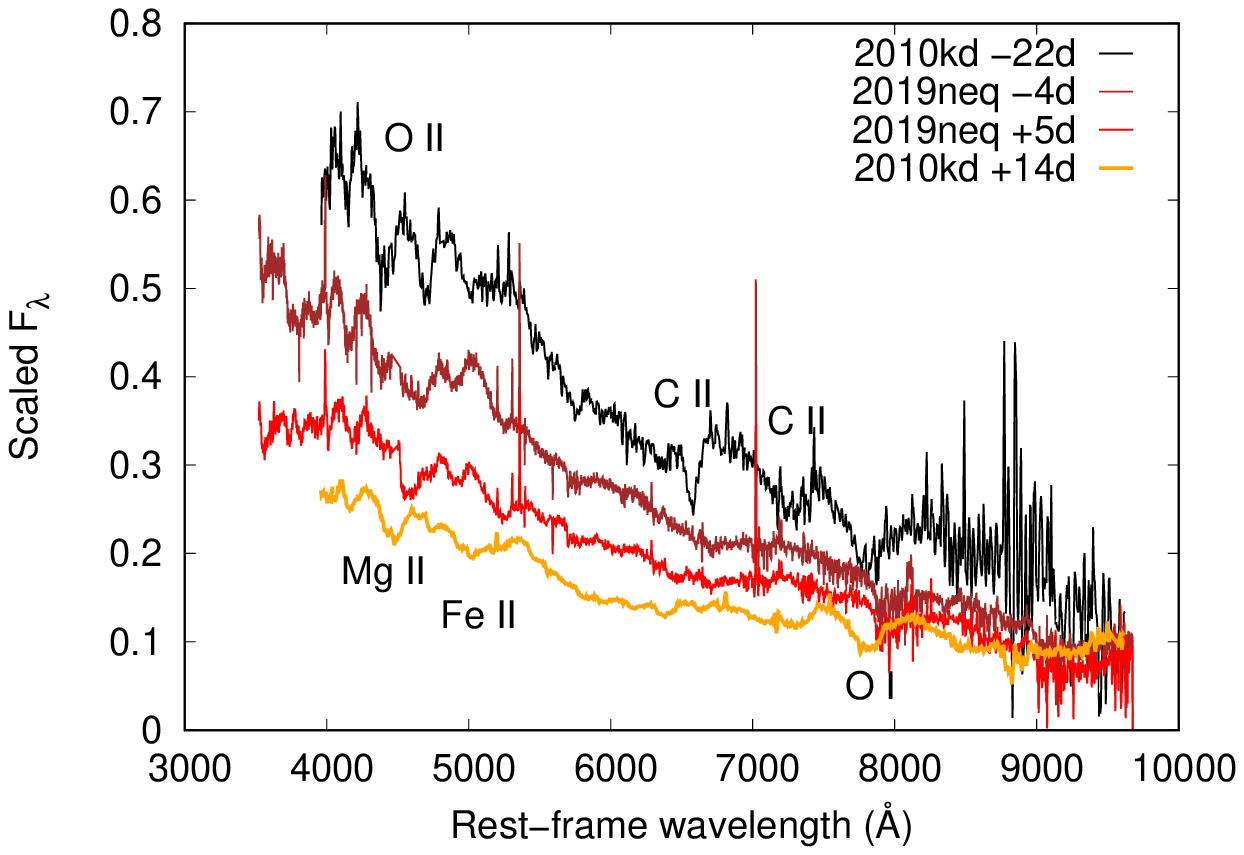}
\includegraphics[width=8cm]{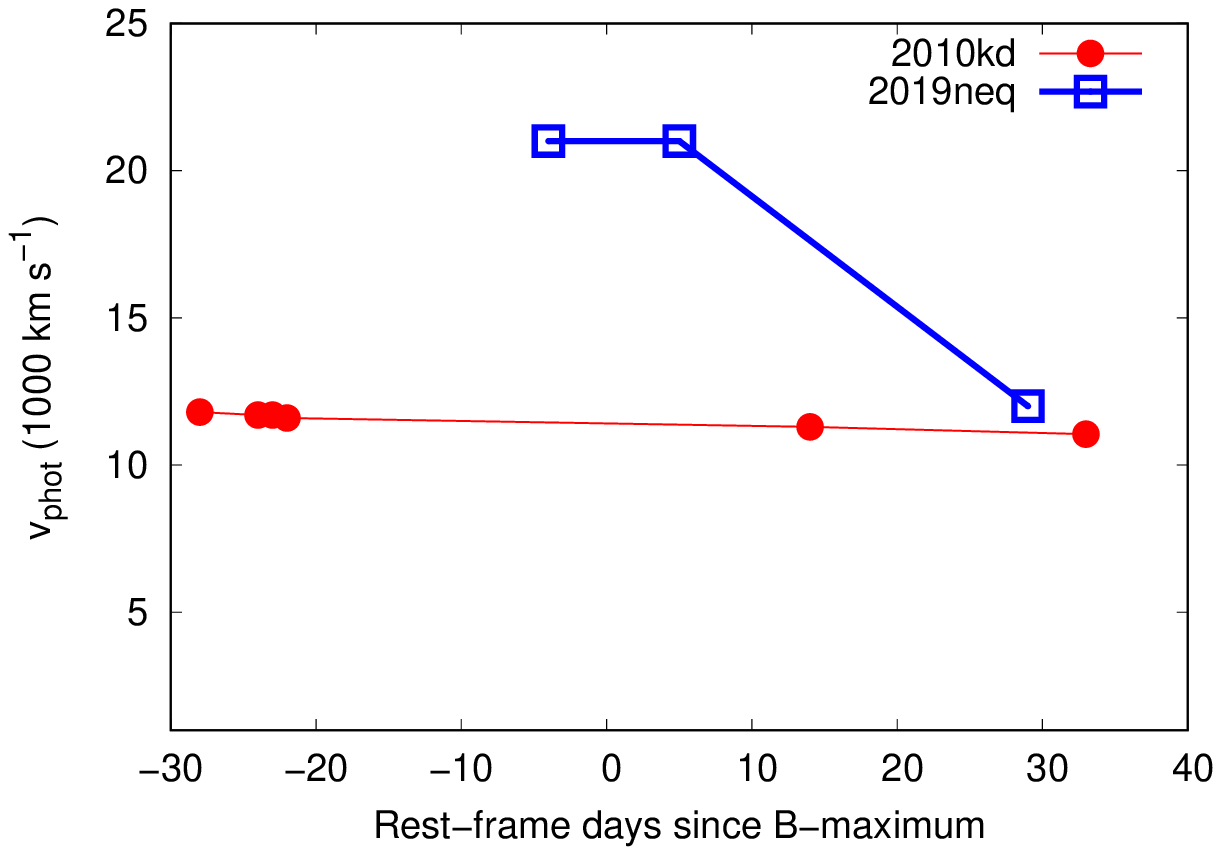}
\includegraphics[width=8cm]{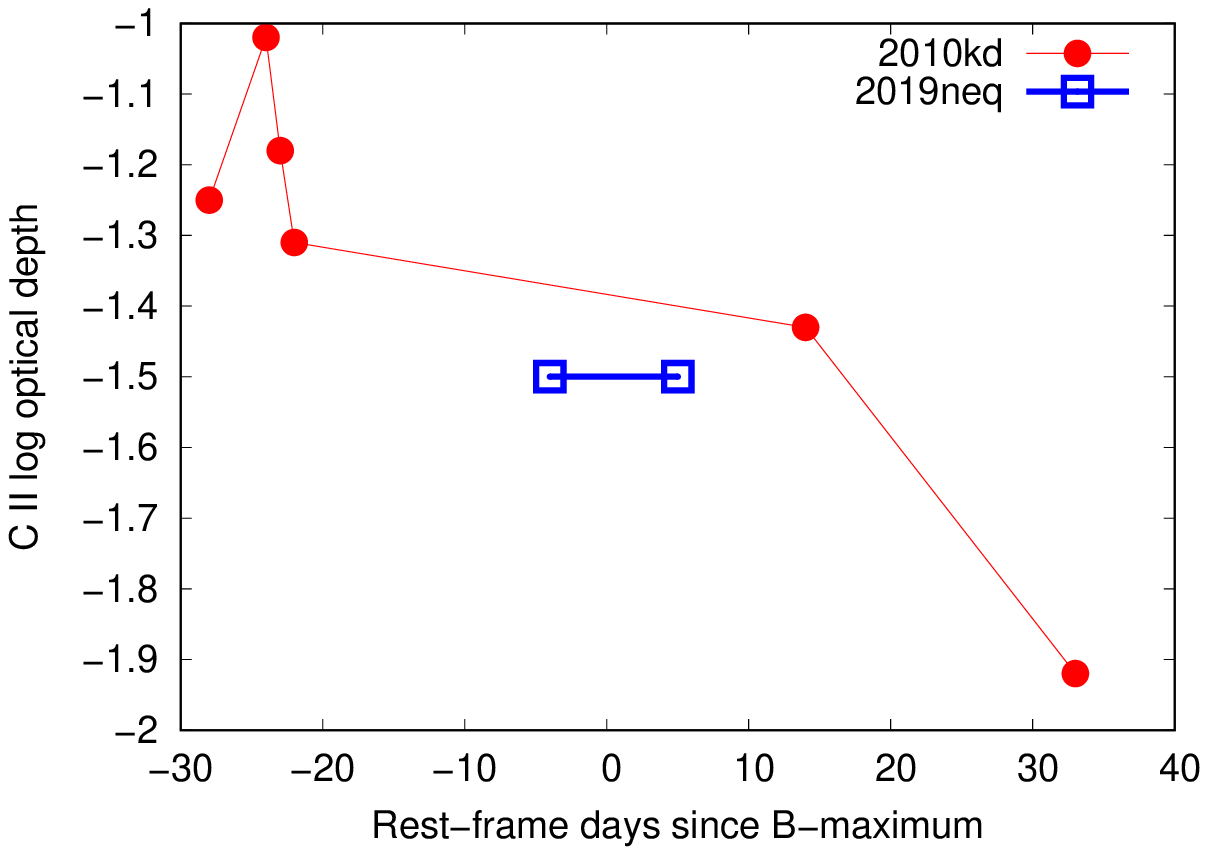}
\includegraphics[width=8cm]{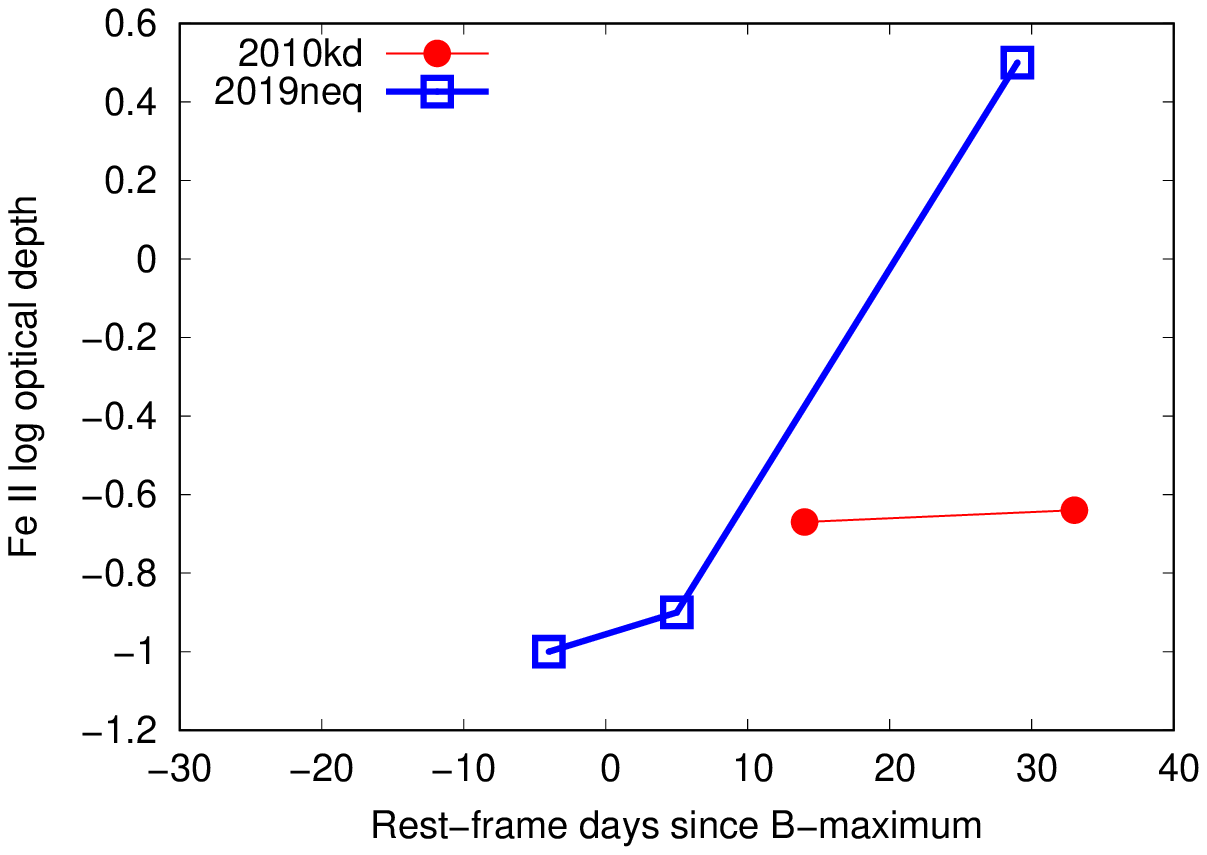}
\caption{Comparison of the spectral evolution of SN~2019neq and SN~2010kd before and shortly after maximum light. {\it Top left panel:} Spectra Doppler-shifted back to zero photospheric velocity, showing the major identified features. {\it Top right panel:} The evolution of the photospheric velocities. {\it Bottom left panel:} Evolution of the optical depth of the C II reference line. {\it Bottom right panel:} the same as the bottom left panel but for the Fe II reference feature.}
\label{fig:spcomp}
\end{center}
\end{figure*}

In this subsection we compare the spectral evolution of the fast SLSN, SN~2019neq, to slowly evolving SLSNe. We selected SN~2010kd as a representative example for the latter. The spectroscopic modeling of SN~2010kd, computed with {\tt SYNAPPS} ({\tt SYN++} coupled with an automated parameter optimization routine), was published recently by \citet{kumar19}.

Figure \ref{fig:spcomp} presents plots for comparing various spectroscopic quantities. Spectra of the two SLSNe taken before or shortly after maximum light are shown in the  top left panel with major features identified by the {\tt SYN++} models. It can be seen that the pre-maximum spectra are globally similar: they are dominated by a hot blue continuum with some (weak) ionized carbon and oxygen features. This remains true for  the early post-maximum phases, even though the decrease of the continuum slope implies a cooling ejecta for both classes. One apparent difference between SN~2010kd and SN~2019neq is the characteristic time-scale of their spectral evolution: the +5d spectrum of SN~2019neq has similar continuum slope to that of SN~2010kd at phase +14d, again, implying that SN~2019neq is a fast evolving SLSN-I (see Section~\ref{fig:moses_vs_neq}).

In the top right panel the evolution of the photospheric velocities are shown as a function of the rest-frame phase since B-band maximum in the case of SN~2010kd and ZTF g'-band maximum in the case of SN~2019neq. It is seen that the velocity of SN~2010kd is nearly constant through the observed epochs, which implies that the outer ejecta remain optically thick up to $\sim +35$ rest-frame days after maximum. On the other hand, SN~2019neq shows a factor of $\sim 2$ higher $v_{\rm phot}$ around maximum light that quickly decreases to $\sim 12000$ km~s$^{-1}$ (a more typical $v_{\rm phot}$ value for SLSNe) by +30d phase \citep{ben}. This fast velocity decline is probably caused by the quick decrease of the density in the outer ejecta, which may suggest a different density profile and somewhat lower ejecta mass for SN~2019neq compared to SN~2010kd. This is consistent with the results of the mass estimates presented in Section \ref{mass} below.  

At B-band maximum, the photospheric velocities imply photospheric radii $r_{\rm phot} \sim 6 \cdot 10^{15}$ cm for SN~2010kd and $r_{\rm phot} \sim 5 \cdot 10^{15}$ cm for SN~2019neq. The similar value of $r_{\rm phot}$ is due to the fact that the larger velocity of SN~2019neq compensates the shorter rise-time from explosion to maximum light. 

In the bottom panels of Figure \ref{fig:spcomp}, the evolution of the  optical depths of C II and Fe II are plotted. 
The optical depth of C II is the same order of magnitude for SN~2010kd and SN~2019neq, if present. Both objects show a swift fall off in $\log \tau$ after the maximum: up to  +30d phase, the $\log \tau$ value of SN~2010kd decreases to $\sim -2$, while C II flux is not detected in the case of SN~2019neq. This behavior is consistent with the observations of other SLSNe, where the carbon features can be found only before or around maximum, and they quickly diminish in post-maximum phases \citep[e.g.][]{inserra18,quimby18}.

On the contrary, the optical depth of Fe II seems to be different for the two objects: SN~2010kd shows nearly constant values after maximum, while the Fe II optical depth of SN~2019neq rises rapidly in this phase. This is related to the strengthening of the Fe II features with decreasing temperature, as seen e.g. in the post-maximum spectra of Type SNe Ia during the ``Fe II-phase'' \citep{branch17}. The Fe II optical depth estimate for the +29d spectrum of SN~2019neq is based on only a single feature, thus it may be overestimated.

\subsection{Lower limits to the ejecta mass }\label{mass}

\begin{table*}
\caption{Estimates for the ejecta mass from the total optical depth.}
\label{tab:gomb}
\centering
\begin{tabular}{ccccc}
\hline
SN & Days from explosion & $r$ ($10^{15}$cm) & $\tau_{tot}$ & $M_{ej}$ (M$_\odot$)  \\
\hline
SN~2010kd & 52.5 & 6.15 & 60.00 & 47.83  \\
SN~2019neq & 31.0 & 5.09 & 42.80 & 23.32 \\
SNe Ia & 18.0 & 1.56  & 28.42 & 1.44\\
\hline
\end{tabular}
\end{table*}

It is possible to give constraints on the ejecta mass 
from the criterion that the total optical depth ($\tau_{tot}$) for the inner, opaque ejecta should be  $\tau_{tot} > 1$ during the photospheric phase. Since $\tau_{tot} \sim \kappa \cdot \rho \cdot r_{\rm phot}$, where $r_{\rm phot}$ can be inferred from the expression of homologous expansion as $r_{\rm phot} = v_{phot} \cdot (t - t_0) / (1+z)$ (where $t_0$ is the explosion date).  Then, we estimate the density from the following formula:
\begin{equation}
  \rho~=~{{\tau_{tot}} \over {\kappa\cdot r_{\rm phot}}} .
    \label{eq:tauteljes}
\end{equation}
The total optical depth below the photosphere around maximum, $\tau_{tot}$, can be inferred from the formulae of \citet{arnett96} \citep[see also][]{branch17} as $\tau_{tot}~\approx~3c/v_{sc}$, where $v_{sc}$ is the scaling velocity of the homologously expanding ejecta that we approximate with $v_{sc} ~= ~v_{phot}$ at maximum light.
We also assume that the total opacity, $\kappa$, inside the opaque SN ejecta can be approximated by the Thompson scattering opacity of a H-poor SN envelope, $\kappa \sim 0.1$ cm$^2$~g$^{-1}$. 

Finally, after getting the density via Eq. \ref{eq:tauteljes}, the total ejecta mass is estimated by assuming a constant density distribution and using the photospheric radius from the homologous expansion: 
\begin{equation}
  M_{ej}~=~{{4 \pi} \over {3}} r_{\rm phot}^3 \cdot  \rho ~=~
 {{4 \pi} \over {3}}  {{v_{phot}^2 \cdot (t-t_0)^2} \over {(1+z)^2} } \cdot {\tau_{tot} \over \kappa}  
    \label{eq:mass}
\end{equation}

The predicted radius, ejecta mass and optical depth values of SN~2019neq and SN~2010kd at maximum light can be found in Table \ref{tab:gomb}. 

Table \ref{tab:gomb} shows 
that the predicted mass limit for the Slow SLSN-I, SN 2010kd ($\sim 48$ $M_\odot$) is more than a factor of 2 higher than the value belonging to the fast evolving one, SN~2019neq ($\sim 23$ $M_\odot$). Since these are only order of magnitude estimates, we cannot draw the conclusion that faster SLSNe possess less ejecta mass than slower SLSNe, but it would be interesting to find a correspondence to the Phillips-relation for normal SNe \citep{nicholl15}.
In order to test this hypothesis, examination of a larger sample of SLSNe is planned. 

As a cross-check, we also compared our optical depth estimates to the inferred $\tau_{tot}$ of a normal Type Ia SN at maximum, derived in the same way as above, using $v_{phot}~=~10000$ km~s$^{-1}$, $(t-t_0) ~/~ (1+z)~=~18$ days, and $M_{ej}~=~1.44 M_\odot$. This gave $\tau_{tot} ~=~28.42$, which is roughly similar to the optical depths of the SLSNe listed in Table \ref{tab:gomb}.  
This suggests that the masses given in Table \ref{tab:gomb} are valid order of magnitude estimates  of the true ejecta masses.

Studying the nebular spectra of SLSNe can result in another constraint on the ejecta mass, since the whole atmosphere of the SN becomes transparent by this phase, revealing the innermost layers of the object. According to  \citet{maurer10},  $\sim$70\% of the ejecta mass of type Ibc SNe comes from oxygen. 
The oxygen mass of SN~2010kd  was recently published by \citet{kumar19} as $\sim$ 20 $M_\odot$. This is consistent with the mass derived from their bolometric light curve modeling, and can be considered as a lower limit to the entire ejecta mass. 
We are planning to obtain nebular spectra of SN 2019neq when it emerges from solar occulusion.

\section{Conclusions} \label{sec:concl}

We present a comparative spectral analysis of the recently discovered Fast SLSN-I 2019neq with the well-observed, Slow SLSN-I 2010kd \citep{kumar19} by modeling their photospheric phase spectra. 

The redshift- and extinction-corrected spectra of SN 2019neq at the 3 observed epochs (-4d, +5d, +29d) were modelled using the {\tt SYN++} code \citep{thomas11}. The photospheric velocity in the first two spectra were roughly constant at 21000 km~s$^{-1}$, then  suddenly dropped to 12000 km~s$^{-1}$ by the epoch of the third observation (+29d), suggesting a very fast velocity evolution \citep{ben}. Over the same period, the photospheric temperature decreased from  15000 K to 12000 K, then to 6000 K.

In the first spectrum of SN~2019neq (-4d), we identified C II, C III, O III, Si III, Si IV, Co III, and Fe II lines. An alternative model containing C II and O II instead of C III and O III was found to describe the W-shaped feature between 4000 and 5000 \AA\ as well as the previous model. This ambiguity is consistent with \citet{hatano99}, who demonstrated that around $T \sim 15000$ K, the optical depths of the pairs of ionization states O II and O III, as well as C II and C III, are similar.

The second spectrum at +5d contains similar elements and ionisation states as the previous epoch, together with newly appearing C I lines. While a photospheric velocity of $v_\mathrm{phot}$ = $16000$ km s$^{-1}$ is suggested from the apparent position of the Fe II minimum, we found that a model with  $v_\mathrm{phot}$ = $21000$ km s$^{-1}$ more accurately fits the observed features.
We found that this wavelength region is dominated by the blending of numerous, weak Fe II lines, and that the observed feature minimum is unlikely to correspond to the line of Fe II $\lambda 5169$.

The spectrum of the third epoch differs from the previous ones regarding both the ion composition and the photospheric velocity. Since the photospheric temperature decreased to 6000 K, the neutral and low-ionized elements began to dominate over the lines of the highly ionized transitions that were present in the earlier spectra. At this epoch we identified O I, Na I, Mg II, Si II 
and  Fe II. 

From the available spectra, it was possible to classify SN~2019neq by comparing its +29d phase spectrum to the +85d spectrum of the slow evolving SN~2010kd. Since the two spectra are quite similar, we concluded SN~2019neq to be a spectroscopically fast evolving SLSN-I. 

Using the optical depths of the reference features for each ion from our SYN++ models, we inferred the local densities of each ion at the three observational epochs,  and thereby reveal the chemical composition of the object.

The comparison of the evolution of the photospheric velocity and the optical depths of strong features (C~II and Fe~II in particular) of SN~2019neq with those of SN~2010kd suggests somewhat different ejecta parameters, such as the density profile and the total mass.

We  also estimated the total ejecta mass from the expected optical depth around maximum light \citep{branch17}, and found $M_{ej} \sim 23$ and $\sim 48$ $M_\odot$ for SN~2019neq and SN~2010kd, respectively.
These are consistent with the mass estimates from light curve modeling (20-40 $M_\odot$) given by \citet{manos13} and \citet{nicholl16}, and exceed the typical SN-Ia ejecta mass by at least one order of magnitude. Furthermore, we found a possible correlation between the ejecta mass and  evolution time scale of SLSNe: faster evolving SLSNe may have  lower ejecta mass. Since this statement is based on a small sample of objects, testing the reliability of this hypothesis  requires many more SLSNe to be modelled using similar methods to that described above.

\acknowledgments

RKT and JV are supported by the project ``Transient Astrophysical Objects" GINOP 2.3.2-15-2016-00033 of the National Research, Development and Innovation Office (NKFIH), Hungary, funded by the European Union.
RKT is also supported by the \'UNKP-19-02 New National Excellence Program of the Ministry for Innovation and Technology.
JCW and BPT are supported in part by NSF grant 1813825. This study is based on observations obtained with the Hobby-Eberly Telescope, which is a joint project of the University of Texas at Austin, the Pennsylvania State University, Ludwig-Maximilians-Universität München, and Georg-August-Universität Göttingen.
The HET is named in honor of its principal benefactors, William P. Hobby and Robert E. Eberly. 
The Low Resolution Spectrograph 2 (LRS2) was developed and funded by the University of Texas at Austin McDonald Observatory and Department of Astronomy and by Pennsylvania State University. We thank the Leibniz-Institut für Astrophysik Potsdam (AIP) and the Institut für Astrophysik Göttingen (IAG) for their contributions to the construction of the integral field units. 

{}

\section{Appendix}

\begin{table*}[h!]
\caption{Best-fit local parameters of the SYN++ photospheric phase models of  SN~2019neq.}
\label{tab:neq_lokparams}
\centering
\begin{tabular}{cccccc}
\hline 
Element & $\log\tau$ & $v_{\rm min}$ & $v_{\rm max}$ & $\sigma$ & $T_{\rm exc}$ \\
     &           & ($10^3$ km~s$^{-1}$) & ($10^3$ km~s$^{-1}$) & ($10^3$ km~s$^{-1}$) & ($10^3$ K) \\
\hline
\hline
MJD 58727 (-4) &&&&&\\
\hline
C II & -1.2 & 21.0 & 50.0 & 5.0 & 15.0 \\
C III & -0.2 & 21.0 & 50.0 & 2.0 & 30.0 \\
O III & 1.0 & 21.0 & 50.0 & 1.0 & 15.0 \\
Si III & 0.2 & 21.0 & 50.0 & 2.0 & 20.0 \\
Si IV & 0.0 & 21.0 & 50.0 & 2.0 & 20.0 \\
Fe II & -1.0 & 21.0 & 50.0 & 2.0 & 15.0 \\
Co III & -0.5 & 21.0 & 50.0 & 2.0 & 20.0\\
\hline
MJD 58727 (-4) Alternative model &&&&&\\
\hline
C II & -1.5 & 21.0 & 50.0 & 1.0 & 15.0 \\
O II & -1.7 & 21.0 & 50.0 & 1.0 & 15.0 \\
Si III & 0.0 & 21.0 & 50.0 & 2.0 & 20.0 \\
Fe II & -1.0 & 21.0 & 50.0 & 1.0 & 15.0 \\
Co III & -0.5 & 21.0 & 50.0 & 2.0 & 20.0 \\
\hline
MJD 58737 (+5) &&&&&\\
\hline
C I & 0.0 & 21.0 & 50.0 & 2.0 & 12.0 \\
C II & -1.5 & 21.0 & 50.0 & 5.0 & 12.0 \\
O III & 0.7 & 21.0 & 50.0 & 1.0 & 12.0 \\
Si III & 0.2 & 21.0 & 50.0 & 2.0 & 12.0 \\
Si IV & -0.3 & 21.0 & 50.0 & 2.0 & 12.0 \\
Fe II &-0.9 & 21.0 & 50.0 & 2.0 & 12.0 \\
Co III & -0.5 & 21.0 & 50.0 & 2.0 & 18.0 \\
\hline
MJD 58763 (+29) &&&&&\\
\hline
O I & 0.0 & 12.0 & 50.0 & 5.0 & 6.0 \\
Na I & -0.2 & 12.0 & 50.0 & 2.0 & 6.0 \\
Mg II & 0.7 & 12.0 & 50.0 & 2.0 & 6,0 \\
Si II & 0.3 & 12.0 & 50.0 & 2.0 & 6.0 \\
Fe II & 0.5 & 12.0 & 50.0 & 2.0 & 11.0 \\
\hline
\end{tabular}
\end{table*}

\begin{table*}[h!]
 \caption{The parameters required to compute the number density of individual species of SN~2019neq for each epoch. }
\label{tab:koncentracio_input_neq}
\centering
\begin{tabular}{ccccccccc}
\hline 
Element & $\log\tau$ & $g$ & $\log(gf)$ & $T$ (K) & $\lambda$ (\AA\ ) & $t_d$ (days) & $z(T)$ & $\chi$ (eV) \\
\hline
\hline
MJD 58727 (-4) &&&&&&& \\
\hline
C II & -1.2 & 6 & 0.77 & 15000 & 4267 & 24.4 & 6.18 & 18.07 \\
C III & -0.2 & 3 & 0.08 & 30000 & 4647 & 24.4 & 1.77 & 29.57 \\
Si III & 0.2 & 5 & 0.18 & 15000 & 4553 & 24.4 & 1.06 & 19.04 \\
Si IV & 0.0 & 2 & 0.20 & 20000 & 4089 & 24.4 & 2.03 & 24.08 \\
Fe II & -1.0 & 10 & -1.40 & 15000 & 5018 & 24.4 & 100.64 & 2.89 \\
Co III & -1.0 & 8 & -2.36 & 20000 & 4433 & 24.4 & 46.64 & 10.41 \\
\hline
MJD 58727 (-4) Alternative &&&&&&&\\
\hline
C II & -1.5 & 6 & 0.77 & 15000 & 4267 & 24.4 & 6.18 & 18.07 \\
Si III & 0.00 & 5 & 0.18 & 20000 & 4553 & 24.4 & 1.21 & 19.04 \\
Fe II & -1.0 & 10 & -1.40 & 15000 & 5018 & 24.4 & 100.64 & 2.89 \\
Co III & -0.5 & 8 & -2.36 & 20000 & 4433 & 24.4 & 46.64 & 10.41 \\
\hline
MJD 58737 (+5) &&&&&&& \\
\hline
C I & 0.0 & 5 & 0.07 & 12000 & 9095 & 33.5 & 10.69 & 7.49 \\
C II & -1.5 & 6 & 0.77 & 12000 & 4267 & 33.5 & 6.04 & 18.07 \\
Si III & 0.2 & 5 & 0.18 & 12000 & 4553 & 33.5 & 1.02 & 19.04 \\
Si IV & -0.3 & 2 & 0.20 & 12000 & 4089 & 33.5 & 2.00 & 24.08 \\
Fe II & -0.9 & 10 & -1.40 & 16000 & 5018 & 33.5 & 108.78 & 2.89 \\
Co III & -0.5 & 8 & -2.36 & 18000 & 4433 & 33.5 & 42.81 & 10.41 \\
\hline
MJD 58763 (+29)  &&&&&&& \\
\hline
O I & 0.00 & 5 & 0.32 & 6000 & 7772 & 57 & 8.95 & 9.16 \\
Na I & -0.2 & 2 & 0.12 & 6000 & 5890 & 57 & 2.19 & 0.00 \\
Mg II & 0.7 & 4 & 0.74 & 6000 & 5184 & 57 & 2.00 & 8.87 \\
Si II & 0.3 & 2 & 0.30 & 6000 & 6347 & 57 & 5.73 & 8.13 \\
Fe II & 0.5 & 6 & -1.4 & 11000 & 5018 & 57 & 72.75 & 2.89 \\
\hline
\end{tabular}
\end{table*}

\begin{table*}[h!]
 
\caption{The inferred values of the number and mass densities of each ionization state in SN~2019neq at all epochs.}
\label{tab:koncentracio_out_neq}
\centering
\begin{tabular}{ccccc}
\hline 
Element & $\log {n_l}$ (cm$^{-3}$) & $\log N$ (cm$^{-3}$) & Mass number & $\log \rho$ (g~cm$^{-3}$) \\
\hline
\hline
MJD 58727 (-4d) &&& \\
\hline
C II & 3.02 & 5.67 & 12 & -17.03 \\
C III & 4.71 & 6.64 & 12 & -16.06 \\
Si III & 4.93 & 7.04 & 28 & -15.29 \\
Si IV & 4.44 & 7.08 & 28 & -15.25 \\
Fe II & 5.62 & 7.04 & 56 & -14.99 \\
Co III & 6.60 & 8.51 & 59 & -13.50 \\
\hline
MJD 58727 (-4) Alternative &&&\\
\hline
C II & 2.72 & 5.37 & 12 & -17.33 \\
Si III & 4.86 & 6.33 & 28 & -16.01 \\
Fe II & 7.10 & 9.01 & 56 & -13.02 \\
Co III & 5.62 & 7.04 & 59 & -14.97 \\
\hline
MJD 58737 (+5) &&& \\
\hline
C I & 4.61 & 6.31 & 12 & -16.39 \\
C II & 2.49 & 5.79 & 12 & -16.91 \\
Si III & 4.70 & 7.48 & 28 & -14.85 \\
Si IV & 3.78 & 8.17 & 28 & -14.16 \\
Fe II & 5.61 & 7.04 & 56 & -14.99 \\
Co III & 6.92 & 8.91 & 59 & -13.09 \\
\hline
MJD 58763 (+29)  &&& \\
\hline
O I & 3.83 & 7.42 & 16 & -15.15 \\
Na I & 3.43 & 3.47 & 22 & -18.97 \\
Mg II & 4.01 & 6.95 & 24 & -15.45 \\
Si II & 3.75 & 7.17 & 28 & -15.16 \\
Fe II & 6.39 & 8.05 & 56 & -13.98 \\
\hline
\end{tabular}
\end{table*}


\begin{thebibliography}{}

\bibitem[Arnett(1996)]{arnett96} Arnett, D.\ 1996, Supernovae and Nucleosynthesis: An Investigation of the History of Matter

\bibitem[Branch \& Wheeler(2017)]{branch17} Branch, D., \& Wheeler, J.~C.\ 2017, Supernova Explosions: Astronomy and Astrophysics Library

\bibitem[Chatzopoulos et al.(2013)]{manos13} Chatzopoulos, E., Wheeler, J.~C., Vinko, J., et al.\ 2013, \apj, 773, 76


\bibitem[Fisher(1999)]{fisher99} Fisher, P.\ 1999, Ph.D. Thesis

\bibitem[Gal-Yam(2012)]{galyam12} Gal-Yam, A.\ 2012, Science, 337, 927

\bibitem[Hatano et al.(1999)]{hatano99} Hatano, K., Branch, D., Fisher, A., et al.\ 1999, \apjs, 121, 233

\bibitem[Inserra et al.(2013)]{inserra13} Inserra, C., Smartt, S.~J., Jerkstrand, A., et al.\ 2013, \apj, 770, 128

\bibitem[Inserra et al.(2018)]{inserra18} Inserra, C., Prajs, S., Gutierrez, C.~P., et al.\ 2018, \apj, 854, 175

\bibitem[Inserra(2019)]{inserra19} Inserra, C.\ 2019, Nature Astronomy, 3, 697

\bibitem[Konyves-Toth et al.(2019)]{atel} Konyves-Toth, R., Vinko, J., Thomas, B. P., Wheeler, J. C.\ 2019, The Astronomer's Telegram,13083, 1

\bibitem[Kumar et al.(2020)]{kumar19} Kumar, Pandej, Konyves-Toth et al., 2020, \apj in press

\bibitem[Maurer \& Mazzali(2010)]{maurer10} Maurer, I., \& Mazzali, P.~A.\ 2010, \mnras, 408, 947

\bibitem[Mazzali et al.(2016)]{mazzali16} Mazzali, P.~A., Sullivan, M., Pian, E., et al.\ 2016, \mnras, 458, 3455

\bibitem[Mulligan et al.(2019)]{mulligan19} Mulligan, B.~W., Zhang, K., \& Wheeler, J.~C.\ 2019, \mnras, 484, 4785


\bibitem[Nicholl et al.(2015)]{nicholl15} Nicholl, M., Smartt, S.~J., Jerkstrand, A., et al.\ 2015, \mnras, 452, 3869

\bibitem[Nicholl et al.(2016)]{nicholl16} Nicholl, M., Berger, E., Smartt, S.~J., et al.\ 2016, \apj, 826, 39

\bibitem[Nicholl et al.(2018)]{nicholl18} Nicholl, M., Blanchard, P.~K., Berger, E., et al.\ 2018, \apjl, 866, L24

\bibitem[Perley et al.(2019)]{perley19} Perley, D.~A., Yan, L., Gal-Yam, A., et al.\ 2019, Transient Name Server AstroNote, 79, 1

\bibitem[Quimby et al.(2007)]{quimby07} Quimby, R.~M., Aldering, G., Wheeler, J.~C., et al.\ 2007, \apjl, 668, L99

\bibitem[Quimby et al.(2018)]{quimby18} Quimby, R.~M., De Cia, A., Gal-Yam, A., et al.\ 2018, \apj, 855, 2

\bibitem[Silverman et al.(2015)]{silver15} Silverman, J.~M., Vink{\'o}, J., Marion, G.~H., et al.\ 2015, \mnras, 451, 1973


\bibitem[Smith et al.(2007)]{smith07} Smith, N., Li, W., Foley, R.~J., et al.\ 2007, \apj, 666, 1116

\bibitem[Thomas et al.(2011)]{thomas11} Thomas, R.~C., Nugent, P.~E., \& Meza, J.~C.\ 2011, \pasp, 123, 237

\bibitem[Thomas et al.(2019)]{atel2} Thomas, B.~P., Konyves-Toth, R., Vinko, J.,  et al.\ 2019, The Astronomer's Telegram, 13184, 1

\bibitem[Thomas et al. (2020)]{ben} Thomas, B.~P., Konyves-Toth, R., Vinko, J., Wheeler, J.~C., in prep

\bibitem[Vinko et al.(2010)]{vinko10} Vinko, J., Zheng, W., Romadan, A., et al.\ 2010, Central Bureau Electronic Telegrams 2556, 1

\end{thebibliography}
\end{document}